\documentstyle[prl,aps,epsf]{revtex}
\draft
\begin{document}
\twocolumn[\hsize\textwidth\columnwidth\hsize\csname @twocolumnfalse\endcsname

\title{Colloidal Dynamics on Disordered Substrates}
\author{C. Reichhardt and C.J. Olson} 
\address{ 
Center for Nonlinear Studies, Theoretical, and Applied Physics Divisions, 
Los Alamos National Laboratory, Los Alamos, NM 87545}

\date{\today}
\maketitle
\begin{abstract}
Using Langevin simulations we examine driven colloids interacting
with quenched disorder.
For weak substrates the colloids form an ordered state and depin 
elastically. 
For increasing substrate strength we find 
a sharp crossover to inhomogeneous depinning and
a substantial increase in the depinning force,
analogous to the peak effect in superconductors. The velocity
versus driving force curve shows criticality at depinning, with 
a change in scaling exponent occuring
at the order to disorder crossover.
Upon application of a sudden pulse of driving force,
pronounced transients appear in the disordered regime
which are due to the formation of long-lived 
colloidal flow channels. 
\end{abstract}
\vspace{-0.1in}
\pacs{PACS numbers: 82.70.Dd,64.60.Ht}
\vspace{-0.3in}

\vskip2pc]
\narrowtext
Colloidal crystals are an  
ideal system in which to study 
the general problem of ordering and 
dynamics in 2D
\cite{Grier,Nm,Peeters,Bechinger,Frey,Lin}, since
the particle size permits direct imaging 
of the particle locations and motion. 
A considerable amount of work has been conducted on
the melting of 2D colloidal crystals in the absence
of a substrate \cite{Grier,Nm,Peeters}. 
In addition, a number of 
experimental and theoretical studies have 
considered colloidal crystallization and melting in 2D systems 
with periodic 1D \cite{Bechinger,Frey} and 
2D substrates \cite{Lin,Ling}, where a rich variety of 
crystalline states can be stabilized.

Colloid crystals  
are also ideal for
studing the ordering and dynamics 
of an elastic media interacting with {\it random} substrates,
a problem that is relevant to a
wide variety of systems such as superconducting vortices,
Wigner crystals, and charge-density waves (CDWs).   
Open issues include
the nature of the dynamical response to applied forces, as well as   
whether an order to disorder transition occurs
as the strength of the random substrate increases.
Recently, LeDoussal and Carpentier have theoretically
investigated the effects
of quenched disorder on the order and melting of
2D lattices and find 
a sharp crossover from the ordered 
Bragg glass (where defects are absent) 
to a disordered or molten state \cite{Giamarchi}. 
They 
predict that the depinning threshold
increases at this crossover due to the softening of the lattice, 
which allows the
particles to better adjust to the substrate. 
A similar mechanism could account for
the peak effect observed in vortex matter in 
superconductors 
\cite{Higgins,Bhattacharya,Henderson,Xiao1,Xiao2,Zeldov}, in 
which the depinning threshold rises dramatically when 
the applied magnetic field is increased. 
In low temperature superconductors, where the vortices 
are fairly stiff so that their behavior can be
considered as effectively 2D, 
recent small angle neutron scattering experiments
have shown that the peak effect
is associated with a sharp disordering or melting 
transition \cite{Ling2}.    

In addition to static properties,
the dynamics of elastic media 
interacting with quenched disorder in 2D 
is a topic of intense study. 
In the disordered region the driven system may
break up into
pinned and flowing regions,  
as observed in 
experiments  \cite{Tonomura} 
and simulations \cite{Dominguez,Vinokur,Marchetti,Olson}
of vortices in
superconductors. Conversely, for weak substrate disorder
the elastic media is defect free and undergoes
elastic depinning,
in which the particles keep the same neighbors as they move. 
Fisher predicted that elastic depinning would show criticality
\cite{Fisher}, and that the velocity vs. force
curves would scale as $v = (f - f_c)^\beta$, where $f_c$ is the depinning
threshold force. This scaling has been studied extensively
in 2D CDW systems where $\beta = 2/3$ \cite{Naryan,Biham,Sethna}.
It is, however, not known whether this exponent occurs in other 
systems undergoing elastic flow.   
Another intriguing dynamical  
phenomena is the pronounced transient behavior exhibited by
vortices under a sudden applied current pulse at magnetic fields
near the peak effect regime \cite{Henderson,Xiao1}. 
Due to surface barrier effects in the 
vortex sample, it is not clear whether these transient 
effects relate to the 
plasticity of the vortex dynamics 
or to contamination of the vortex lattice by disorder from 
the sample edges \cite{Zeldov}.  
Recently, Pertsinidis and Ling \cite{Ling} have 
studied colloids in 2D driven
by an electric field and interacting with a disordered substrate. 
They observe plastic depinning with filamentary or river-like 
flow of colloids and a velocity-force curve scaling with
$\beta = 2.2$. Under a
pulsed drive the system 
shows very long time transients that fit to a 
stretched exponential. 

Motivated by the recent colloidal experiments 
as well as the pulse drive experiments in vortex matter, 
we have conducted Langevin simulations 
of colloidal particles interacting via
a Yukawa potential in 2D systems with random disorder.    
In simulation,
the strength of the disorder can be carefully tuned, which is
difficult to achieve in experiments. In addition,  
the initial conditions of the colloidal arrangements are easily controlled, 
whereas in experiments, 
defects generated in the colloidal lattice
during preparation may become frozen in by the disorder.   

Our simulations show that
for weak substrates the colloids form an 
ordered triangular array 
which depins elastically without the generation of defects. 
For increased substrate strength,
there is a sharp crossover to a disordered phase where the colloids 
depin plastically  
into riverlike structures.
This crossover is accompanied by a sharp increase
in the depinning threshold, analogous to the 
peak effect phenomena in superconductors. 
We 

\begin{figure}
\center{
\epsfxsize=3.5in
\epsfbox{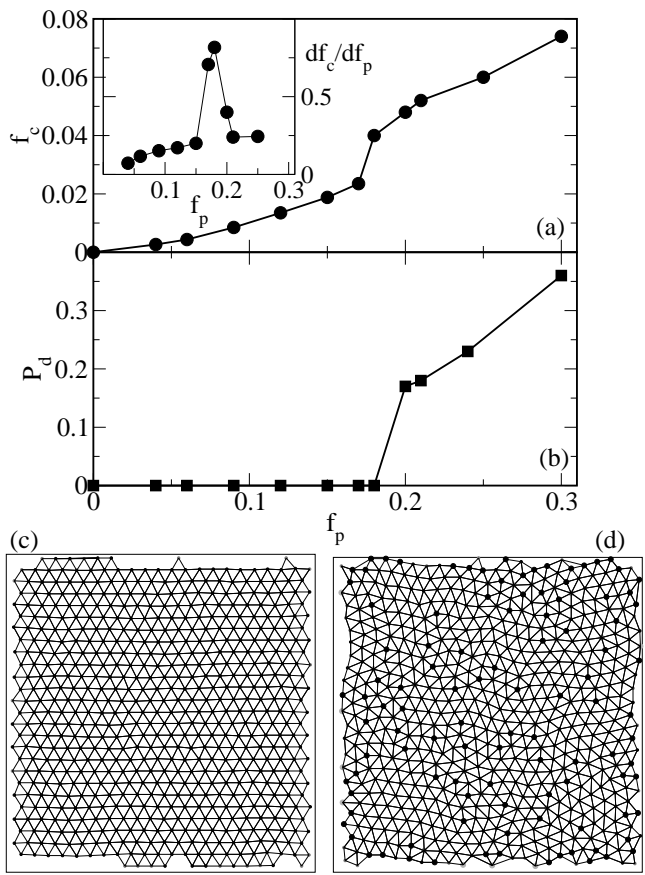}}
\caption{
(a) Depinning force $f_{c}$ vs pinning strength $f_{p}$.
Inset: Corresponding derivative $df_{c}/df_{p}$. 
(b) Percentage of non-sixfold coordinated colloids, $P_{d}$. 
(c) Delaunay triangulation of
colloid positions at depinning for $f_{p} = 0.12$;
(d) $f_{p} = 0.25$.  Filled circles indicate defect sites of
particles which are not sixfold coordinated.
}
\end{figure}

\hspace{-13pt}
find scaling of the velocity vs. applied drive 
with an exponent of $\beta = 0.67$ in the
elastic regime, in agreement with studies in  2D CDW's. 
In the plastic regime we find $\beta = 1.97$, close to the value observed
in experiments \cite{Ling}. 
In the disordered region,
long time transients that fit to a stretched exponential occur 
in response to a sudden applied 
drive pulse, as also observed in the colloidal and
vortex experiments.  

The colloids are simulated using
Langevin dynamics in 2D 
\cite{Nm,Peeters}.
The colloids interact via a Yukawa or screened Coulomb
interaction potential 
$V(r_{ij}) = (Q^2/|{\bf r}_{i} - {\bf r}_{j}|)
\exp(-\kappa|{\bf r}_{i} - {\bf r}_{j}|)$. Here $Q$ is the charge of the particles, $1/\kappa$ is the screening length, and ${\bf r}_{i (j)}$ 
is the position of
particle $i (j)$.
The length of the system is measured in units of the lattice constant 
$a_{0}$ and 
a screening length of $\kappa = 2/a_{0}$ is used.
The quenched disorder on the 
substrate is modeled as randomly placed parabolic traps with 
radius $r_{p} < a_{0}$ and a maximum force $f_{p}$.   
This same type of model for pinning has been used previously to model quenched
disorder in
superconducting vortex systems \cite{Olson}. Other types of pinning
potentials used in vortex simulations produce results similar
to the parabolic pinning \cite{Dominguez,Vinokur,Marchetti}.    
The equation of motion for colloid $i$ is
\begin{equation}
\frac{d {\bf r}_{i}}{dt} = {\bf f}_{ij} + {\bf f}_{p} + {\bf f}_{T} + 
{\bf f}_{d}
\end{equation}
Here ${\bf f}_{ij} = \sum_{j \neq i}^{N_{c}}\nabla_i V(r_ij)$ 
is the interaction force from the other colloids, ${\bf f}_{p}$ is the
pinning force, ${\bf f}_{T}$ is a randomly fluctuating force due to thermal
kicks, and ${\bf f}_{d}$ is the force due to an applied driving field.   
We start the system at a temperature above melting as determined from
the diffusion, and gradually cool below $T_{m}$ to $T/T_{m} = 0.4$ for most
of the data presented here. To measure velocity $v$ vs force curves,
care must taken to average over substantial amounts of time to   
avoid transient effects. 
The driving force is increased from zero by small increments and
the velocity is averaged for $5\times10^{4}$ time steps at each
increment, with typical simulations
running for $10^7$ time steps. In this model we do not take
into account hydrodynamic effects or long-range attractions between
colloids.

In Fig.~1(a) we show the depinning force $f_{c}$
vs. substrate strength $f_{p}$ from a
series
of simulations.
For $f_{p} < 0.18$ the depinning force 
increases as a power law, $f_{c} \propto f_{p}^{-1.9 \pm 0.1}$. 
To compare the depinning force to the order in the
system, in Fig.~1(b)  we show the  
the percentage of defects or non-six fold coordinated particles 
$P_{d}$ 
as calculated from a Delaunay
triangulation. This measure indicates that the colloidal crystal is in
an ordered state ($P_{d} = 0.0$) for $f_{p} < 0.18$, 
and that there is a crossover to a 
disordered state ($P_{d} \neq 0$) at $f_{p}=0.18$.
In Fig.~1(c) we show a
representative Delaunay triangulation for the ordered state 
where there are no defects but
small distortions in the particle positions can be seen,
and in Fig~1(d) we show the disordered state
where defects are present. 
The crossover to the disordered state
coincides with a rapid {\it increase} in the depinning force
as seen in Fig.~1(a) and in the inset of Fig.~1(a) which shows
a peak in $df_{c}/df_{p}$ at the crossover. 
This behavior 
is consistent with the recent experiments
in superconductors which find that 
at the peak effect there is an increase in the pinning 
with a simultaneous disordering of the lattice \cite{Ling2}. 
For $f_{p} > 0.2$, the 
depinning
scales as $f_c \propto f_p$, as expected for the single particle
pinning regime. The sudden increase in the depinning force results from
the fact that 
the defected colloid lattice is much softer than the ordered lattice,
allowing the colloids to adjust their positions 
to accommodate to the optimal pinning sites.       
We have also investigated this transition for different colloidal
densities and disorder strengths. 
For 
increasing $T$ the order to disorder transition is shifted to 
lower values of $f_{p}$. We have also investigated finite size effects 
for increasing system sizes, and find that the
order-disorder crossover shifts only a small amount before saturating,
while the sharpness of the transition persists with increased system
size. 

\begin{figure}
\center{
\epsfxsize=3.5in
\epsfbox{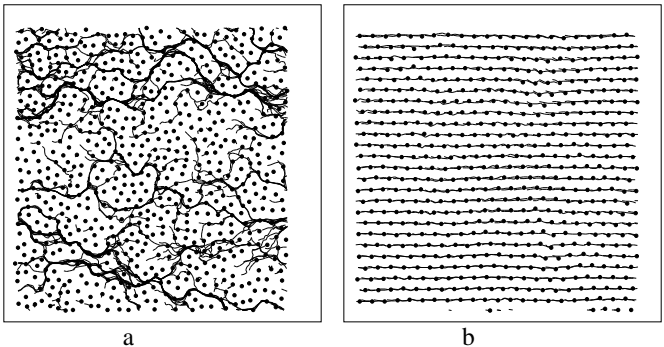}}
\caption{
Colloid positions (black dots) and trajectories (lines) 
for (a) plastic flow regime ($f_{p} = 0.25$), and
(b) elastic flow regime ($f_{p} = 0.12$).
}
\end{figure}

It is beyond the scope of this paper to determine whether the
order to disorder crossover is a first order transition. Although
the sharpness suggests a possible first order transition,
Carpentier and LeDoussal show that for 2D systems with quenched disorder,
a sharp disordering crossover, rather than transition, occurs 
\cite{Giamarchi}. 
In addition, a first order transition is not expected
since the Bragg
glass in 2D has been shown to have dislocations on large scales
at all temperatures. 
The distance 
between these dislocations can be arbitrarily large \cite{Giam}.   

In Fig.~2 we show that 
the order-disorder crossover coincides with the onset of 
plastic flow above depinning. In Fig.~2(b) the 
elastic colloid flow is shown for
$f_{p} = 0.12$  
above depinning ($f_{d}/f_{c} = 1.1$). 
Here 
each colloid keeps the same neighbors as it moves.  In Fig.~2(a) the
inhomogeneous or plastic colloidal flow is shown for 
$f_{d}/f_{c} = 1.1$ for $f_p = 0.25$. Here, only
a portion of the
colloids are moving at any one time, and the motion 
occurs in channels or rivers
between pinned regions.
The colloid velocities show a bimodal distribution in this regime,
split between the stationary and moving colloids.
In addition, the channels seen in Fig.~2(a) are not static but
change over time, so that any one colloid is only temporarily trapped
in a pinning site.  
These features of the plastic flow 
are in agreement with observations in
colloidal experiments \cite{Ling} and 
in vortex simulations 
of the strongly pinned regime \cite{Dominguez,Vinokur,Marchetti,Olson}.

In order to correlate the different types of flow observed in Fig.~2 with
properties of bulk measurements, we show in 
Fig.~3 the scaling of the velocity vs driving force. For
elastic depinning in the 
ordered regime [Fig.~3(a)], the velocity vs force curves fit
well to $v = (f_{d} -f_{c})^\beta$ with $\beta = 0.66 \pm 0.02$, as
illustrated in Fig.~3(b).
These results
are in good agreement with theoretical predictions \cite{Naryan} 
and simulation
results \cite{Biham,Sethna} 
for CDW's in 2D where the depinning is elastic. 
In contrast, in driven 2D vortex matter, 
Higgins and Bhattacharaya \cite{Bhattacharya} 
found an exponent of $\beta = 1.2$ below the peak
effect where elastic flow is expected to occur.
This may be due to the effects of surface barriers
disordering the lattice.
In Fig.~3(c,d) 

\begin{figure}
\center{
\epsfxsize=3.5in
\epsfbox{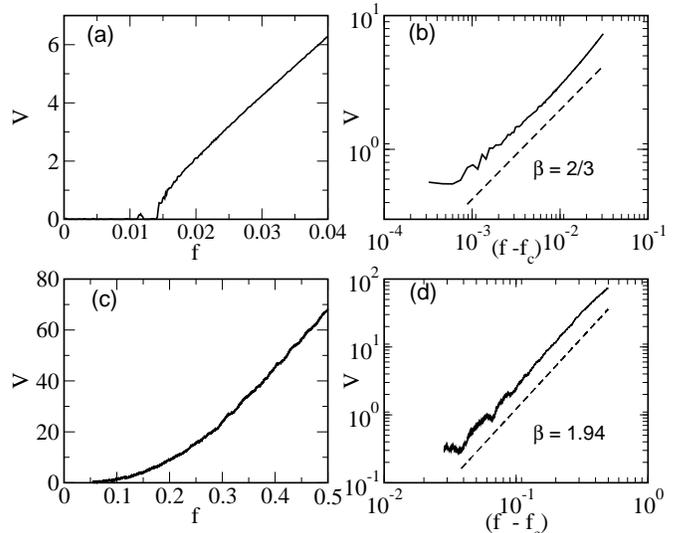}}
\caption{
Velocity $v$ vs applied drive $f_{d}$ for (a) elastic regime
$f_{p} = 0.08$. (b) Log-log plot of $v$ vs $(f_{d} - f_{c})$ from
(a); line indicates fit to $\beta =  2/3$. 
(c) $v$ vs $f_{d}$ for plastic depinning $f_{p} = 0.25$.
(d) Log-log plot of $v$ vs $(f_{d} - f_{c})$ from (c); line
indicates fit to $\beta = 2.2$    
}
\end{figure}

\hspace{-13pt}
the $v-f_{d}$ scaling for the plastic regime 
is presented, 
with
$\beta = 1.94 \pm 0.03$, close to the value of $2.2$ found in
the colloid experiments \cite{Ling}. 
For larger system sizes, we find that
the scaling region is expanded but the exponent is unchanged.
Other studies in the plastic flow regime found 
$\beta = 5/3$ for electron flow in metallic dots \cite{Wingreen} and 
$\beta = 2.22$ for
vortex flow in Josephson-junction arrays \cite{Domg}. 

In Fig.~4 we show the 
response of colloids prepared in an ordered state to
the application of a sudden pulse of driving force
of different strengths in the plastic flow regime.
Since the pulse strength is chosen to be below the depinning
threshold value $f_c$,
the initial colloid velocity is high, and then
gradually decreases. 
We find that a simple functional form cannot be fit to the curves.  
Instead, we use a stretched exponential fit as performed in experiments
\cite{Ling}: 
$v(t) = v_{0}\exp(-(t/t_{0})^\alpha) + v_{1}$.  
The values of $t_{0}$ and $\alpha$ depend on the magnitude of the
drive. For the parameters investigated here, $\alpha$ 
falls between $0.08$ and $0.4$,
in agreement with the values found in experiment. 
A similar stretched exponential decay was also found
in vortex matter for the transient response 
to pulses \cite{Xiao1}.
We find that in the long time limit, the colloid flow occurs only
through a few long-lived channels. 
In the elastic regime,
the decay of $v$ is much faster and fits 
to an initial pure exponential with the velocities
going to zero. 
We note
that in the elastic regime the colloids move less than a lattice constant
after a pulse is applied, whereas
in the plastic regime, colloids in the moving channels can move
the entire length of the system.   
For increased system sizes, the transient times are enhanced in the plastic
flow regime but are unchanged in the elastic regime.  
The long 

\begin{figure}
\center{
\epsfxsize=3.5in
\epsfbox{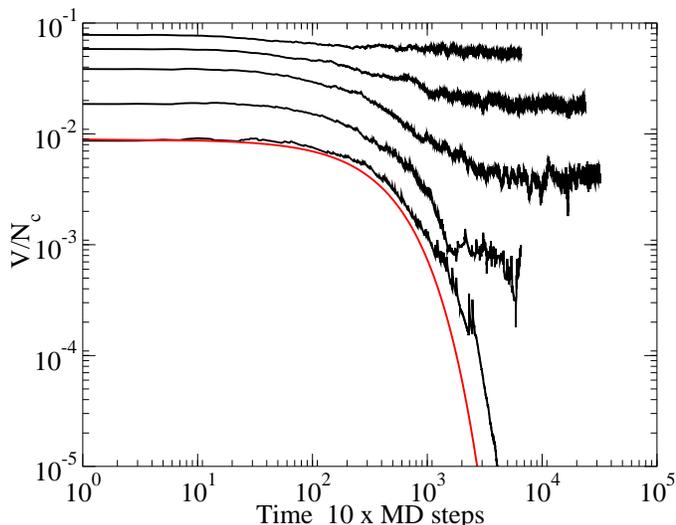}}
\caption{  
Average transient velocity per colloid $V/N_{c}$ vs time in the disordered
region for $f_{p} = 0.25$ at a sudden applied drive of 
(from top to bottom)
$f_{d}/f_{c} = 1.2$, $1.0$, $0.8$, $0.6$, and $0.3$. A stretched 
exponential $A \ \exp (-(t/t_{0})^\alpha) + v_{1}$ can be fit to all the
curves. Bottom curve: a stretched exponential fit for 
the $f_{d}/f_{c}=0.3$ case, with 
$V/N_{c} = 0.0089 \exp (-(t/400)^{1}) + 0$. 
}
\end{figure}

\hspace{-13pt}
lived transients in the plastic regime are responsible
for the very slow velocity-force sweep necessary to 
measure 
an accurate depinning threshold.  
This sweep-rate dependence is also consistent with the experimentally 
observed 
sweep rate dependent critical currents in the peak regime
\cite{Xiao2}, where slow rates produce larger measured critical currents.  

To summarize, we investigated
the behavior of 2D colloids interacting with random disorder
using Langevin simulations. 
For weak disorder the colloids form an ordered lattice which depins
elastically and shows critical
scaling in the velocity vs force curves, with 
$\beta = 0.67$, in agreement with studies of 2D CDW's. For increasing disorder
strength, we find a sharp crossover to a 
disordered state, accompanied by 
a sharp increase in the depinning force, analogous to the 
peak effect observed for
vortex matter in superconductors. In the disordered region, the
colloids depin inhomogeneously into
fluctuating channels and the $v-f$ curve scaling gives 
$\beta = 1.97$, in agreement with experiments.
In the disordered 
flow regime, pronounced transients occur in response to
a sudden pulse, with the
late time dynamics determined by a few long lived channels.    

Acknowledgments. We thank S. Ling and A. Pertsinidis for sharing their
data before publication and for useful discussions. 
We also thank
S.~Bhattacharya, A.~Bishop,  D. Fisher, D.G.~Grier, M.B.~Hastings 
and A.A.~Middleton
for helpful comments. 
This work was supported by the US Department of Energy under
contract W-7405-ENG-36.

\end{document}